\documentclass[twocolumn,prd,preprintnumbers,amsmath,amssymb,superscriptaddress, nofootinbib]{revtex4} 
\usepackage{hyperref}
\usepackage{graphicx}
\usepackage{mathrsfs} 
\usepackage{amsmath, amsthm, amssymb}

\newcommand{\be}{\begin{equation}}
\newcommand{\ee}{\end{equation}}
\newcommand{\beq}{\begin{equation}}
\newcommand{\eeq}{\end{equation}}
\newcommand{\bea}{\begin{eqnarray}}
\newcommand{\eea}{\end{eqnarray}}

\newcommand{\gsim}{\lower.7ex\hbox{$\;\stackrel{\textstyle>}{\sim}\;$}}
\newcommand{\lsim}{\lower.7ex\hbox{$\;\stackrel{\textstyle<}{\sim}\;$}}

\begin{document}

\title{Quantifying the tension between the Higgs mass and $(g-2)_\mu$
  in the CMSSM}

\author{Maria Eugenia Cabrera}
\email[E-mail: ]{maria.cabrera@uam.es}
\affiliation{Instituto de F\'isica Te\'orica, IFT-UAM/CSIC\\
             U.A.M., Cantoblanco\\
             28049 Madrid, Spain}

\author{J. Alberto Casas}
\email[E-mail: ]{alberto.casas@uam.es}
\affiliation{Instituto de F\'isica Te\'orica, IFT-UAM/CSIC\\
             U.A.M., Cantoblanco\\
             28049 Madrid, Spain}

\author{Roberto Ruiz de Austri}
\email[E-mail: ]{rruiz@ific.uv.es}
\affiliation{Instituto de F\'isica Corpuscular, IFIC-UV/CSIC \\
        Valencia, Spain }

\author{Roberto Trotta}
\email[E-mail: ]{r.trotta@imperial.ac.uk}
\affiliation{Astrophysics Group, Imperial College London, Blackett Laboratory \\
       Prince Consort Rd, London SW7 2AZ, UK }

\begin{abstract}
{\small

  Supersymmetry has been often invoked as the new physics that might
  reconcile the experimental muon magnetic anomaly, $a_\mu$, with the
  theoretical prediction (basing the computation of the hadronic
  contribution on $e^+e^-$ data). However, in the context of the
  CMSSM, the required supersymmetric contributions (which grow with
  decreasing supersymmetric masses) are in potential tension with a
  possibly large Higgs mass (which requires large stop masses). In the
  limit of very large $m_h$ supersymmetry gets decoupled, and the
  CMSSM must show the same discrepancy as the SM with $a_\mu$. But it
  is much less clear for which size of $m_h$ does the tension start to
  be unbearable. In this paper, we quantify this tension with the help
  of Bayesian techniques. We find that for $m_h\geq 125$ GeV the
  maximum level of discrepancy given current data ($\sim 3.2\ \sigma$)
  is already achieved. Requiring less than 3~$\sigma$ discrepancy,
  implies $m_h \lsim 120$ GeV. For a larger Higgs mass we should give
  up either the CMSSM model or the computation of $a_\mu$ based on
  $e^+e^-$; or accept living with such inconsistency.}
\end{abstract}

\keywords{Supersymmetry phenomenology, Higgs physics, anomalous
  magnetic moment, Bayesian statistics}
\preprint{IFT-UAM/CSIC-10-84}

\maketitle


\section{Introduction}
\label{Introduction}

The magnetic anomaly of the muon, $a_\mu= \frac{1}{2}(g-2)_\mu$ has been a classical and powerful test for new physics. As it is known, the present experimental value and some of the theoretical determinations of $a_\mu$ show a remarkable discrepancy, suggesting physics beyond the Standard Model (SM) to account it. However, the situation is still uncertain, due essentially to inconsistencies between alternative determinations of the contribution coming from the hadronic vacuum-polarization diagram, say $\delta_{\rm had}^{\rm SM}a_\mu$. 

This contribution can be expressed in terms of the total hadronic cross section $e^+ e^-\rightarrow$ had. Using direct experimental data for the latter, one obtains a final result for $a_\mu$, which is at more than 3~$\sigma$ from the current experimental determination 
\textbf{\cite{Davier:2009zi}}, namely 
\bea
\label{Discr}
\delta a_\mu&=&a_\mu^{\rm exp}-a_\mu^{\rm SM} = 25.5\pm 8.0\ \times 10^{-10}
\eea
(the quoted error bars are 1~$\sigma$). This discrepancy has been often claimed as a signal of new physics. Obviously, if one accepts this point of view, the discrepancy should be cured by contributions from physics beyond the SM.

Admittedly, such claims are too strong. We are quite aware of past experimental observables in apparent disagreement with the SM prediction, which have eventually converged with it. This has occured due to both experimental and theoretical subtleties that sometimes had not been fully understood or taken into account. As a matter of fact, the experimental $e^+e^-\rightarrow$ {\em had} cross section exhibits some inconsistencies between different groups of experimental data. Using only BABAR data the discrepancy reduces to 2.4~$\sigma$, while without it the discrepancy becomes 3.7~$\sigma$, \cite{Davier:2009zi}. The inconsistency is specially notorious if one considers hadronic $\tau$ decay data, which are theoretically related to the $e^+e^-\rightarrow$ {\em had} cross section. Using just $\tau$-data the disagreement becomes 1.9~$\sigma$, \cite{Davier:2009zi}, \cite{Davier:2009ag}. Although the more direct $e^+e^-$ data are usually preferred to evaluate $a_\mu^{\rm SM}$, these inconsistencies are warning us to be cautious about the actual uncertainties involved in the determination of $a_\mu^{\rm SM}$.

If one takes the discrepancy between theory and experiment shown in eq.(\ref{Discr}) as a working hypothesis, one has to consider possible candidates of new physics able to provide the missing contribution to reproduce $a_\mu^{\rm exp}$. The Minimal Supersymmetric Standard Model (MSSM) is then a natural option. We will consider here the simplest and most extensively analyzed version of the MSSM, namely the so-called constrained MSSM (CMSSM), in which the soft parameters are assumed universal at a high scale ($M_X$), where the supersymmetry (SUSY) breaking is transmitted to the observable sector, as happens e.g. in the gravity-mediated SUSY breaking scenario. Hence, our parameter space is defined by the following parameters:
\bea
\label{MSSMparameters}
\{\theta \}\ =\ \{m,M,A,B,\mu,s\}
\ .
\eea
Here $m$, $M$ and $A$ are the universal scalar mass, gaugino mass and trilinear scalar coupling; $B$ is the bilinear scalar coupling; $\mu$ is the usual Higgs mass term in the superpotential; and $s$ stands for the SM-like parameters of the MSSM, i.e. essentially
gauge and Yukawa couplings. All these initial parameters are understood to be defined at $M_X$.

The main supersymmetric (CMSSM) contributions to $a_\mu$ come from 1-loop diagrams with chargino-sneutrino and neutralino-smuon exchange \cite{Stockinger:2006zn}. In general, these contributions, say $\delta^{\rm MSSM}a_\mu $, are larger for smaller supersymmetric masses and can be just of the right magnitude to reconcile theory and experiment (thus constraining the CMSSM parameter space).

In section \ref{higgs:gm2} we show the potential tension between the requirement of suitable SUSY contributions to the muon anomaly and a possibly large Higgs mass. In section \ref{QuantifiyngTheTension} we quantify such tension as a function of $m_h$, with the help of Bayesian techniques.
In section \ref{pdfs} we show how the probability distributions of the most relevant parameters (universal scalar and gaugino masses, and $\tan\beta$) change with increasing $m_h$. Finally, in section 5 we present our conclusions.

\section{Higgs mass vs. g-2}
\label{higgs:gm2}

It is well known that in the MSSM the tree-level Higgs mass is bounded from above by $M_Z$, so radiative corrections (which grow logarithmically with the stop masses) are needed to reconcile the theoretical predictions with the present experimental lower bound, $m_h > 114.4$ GeV (SM-like Higgs). Roughly speaking, a Higgs mass above 130 GeV requires supersymmetric masses above 1 TeV. In this regime one can expect SUSY to be decoupled, so that the prediction for $a_\mu$ becomes close to $a_\mu^{\rm SM}$. Hence, a large Higgs mass in the MSSM would necessarily amounts to a  $\ > 3\ \sigma$ discrepancy between the experimental and the theoretical values of $a_\mu$ (evaluated via $e^+e^-\rightarrow$ had). 

The main goal of this paper is to quantify the tension between $m_h$ and $a_\mu$ in the context of the CMSSM. This is useful since it allows to put an educated upper bound on the Higgs mass, which will depend on the discrepancy one is ready to tolerate. Conversely, it tells us from which minimum value of $m_h^{\rm exp }$ we will have to give up either the CMSSM assumption or the theoretical evaluation of $a_\mu$ via $e^+e^-\rightarrow$ had (with the quoted uncertainties).

For the sake of the discussion, we will give now some approximate analytical expressions for $m_h$ and $\delta a_\mu^{\rm MSSM}$. In the MSSM the tree-level squared Higgs mass plus the one-loop leading logarithmic contribution is given by 
\bea
\label{mhMSSM}
m_h^2&\simeq& M_Z^2 \cos^2 2 \beta \nonumber \\
&&+ {3 m_t^4 \over 2\pi^2 v^2}
\left[\log{m_{\tilde t}^2\over m_t^2} + \frac{X_t^2}{M_S^2}\left(1-\frac{X_t^2}{12M_S^2}\right) \right]\ \\
&&+ \ \cdots \nonumber
\eea
Here $\tan\beta$ is the ratio of the expectation values of the two MSSM Higss fields, $\tan \beta\ \equiv\ \langle H_u\rangle/\langle H_d \rangle$; $m_t$ is the (running) top mass and $m_{\tilde t}$ is the geometrical average of the stop masses. Besides, 
\bea
\label{Xt}
X_t\equiv A_t + \mu \cot \beta,
\eea
where $A_t$ is the top trilinear scalar coupling, and $M_S^2$ is the arithmetical average of the squared stop masses. All the quantities in eqs.(\ref{mhMSSM}), (\ref{Xt}) are understood at low energy (for more details see e.g. ref.\cite{Ellis:1990nz,Ellis:1991zd,Okada:1990vk,Okada:1990gg,Haber:1990aw,Barbieri:1990ja,Casas:1994us}). Subdominant terms not written in eq.(\ref{mhMSSM}) can be important for a precise determination of $m_h$, and we have included them in the numerical analysis. The previous equations tell us how $m_h$ grows with increasing supersymmetric masses and also with increasing $\tan\beta$. Besides, the contribution associated to the stop mixing (second term within the square brackets in eq.(\ref{mhMSSM})) is maximal at $X_t=\sqrt{6}M_S$. 

On the other hand, as mentioned above, the supersymmetric contribution to the muon anomaly, $\delta^{\rm SUSY}a_\mu$, arises mainly from 1-loop diagrams with chargino-sneutrino and neutralino-smuon exchange. This contribution increases with increasing $\tan\beta$ and decreasing supersymmetric masses. See refs.\cite{Degrassi:1998es,Heinemeyer:2003dq,Heinemeyer:2004yq,Marchetti:2008hw,vonWeitershausen:2010zr}. 

 Although the analytical expressions are complicated, one can get an intuitive idea of the parametric dependence by considering the extreme case where the masses of all supersymmetric particles are degenerate at low energy\footnote{This limit is often used because of the simplification of the formulae it implies. However, it is unachievable in the CMSSM.}: $M_1=M_2=\mu
=m_{\tilde \mu L}=m_{\tilde \mu R}=m_{\tilde \nu}\equiv M_{\rm SUSY}$ . Then \cite{Moroi:1995yh},
\bea
\label{g-2}
\delta^{\rm SUSY}a_\mu\simeq \frac{1}{32\pi^2}\frac{m_\mu^2}{M_{\rm SUSY}^2} g_2^2\tan\beta \;{\rm sign}(M_2\mu).
\eea
Examining the approximate expressions (\ref{mhMSSM}) and (\ref{g-2}), it is clear that a large $m_h$ and a large $\delta^{\rm SUSY}a_\mu$ will be more easily obtainable (and thus compatible) for larger $\tan\beta$. On the contrary, the larger the supersymmetric masses the larger $m_h$ but the smaller $\delta^{\rm SUSY}a_\mu$, and this is the origin of the potential tension.

However, it is difficult from the previous expressions (or the more sophisticated ones) to conclude for which size of $m_h$ does the tension start to be unbearable. The reason is that a particular value of the Higgs mass, say $m_h = 120$ GeV, can be achieved through eq.(\ref{mhMSSM}) with different combinations of $\tan\beta$, stop masses and $X_t$. Besides, there are many ways, i.e. very different regions in the MSSM parameter space, in which these quantities can have similar {\em low-energy} values. Still, the corresponding contribution $\delta^{\rm SUSY}a_\mu$ can change significatively from one region to another. Unless one performs a complete scan of the parameter space one cannot conclude that the required value of $\delta^{\rm SUSY}a_\mu$ is unattainable for $m_h=120$ GeV. On the other hand, if it is attainable, but only in an extremely tiny portion of the parameter space, this implies a tension between the two observables since 
the consistency between $m_h$ and $a_\mu$ requires a severe fine-tuning. 
And it is possible, in principle, to quantify such tension.

In the analysis we have included two-loop leading corrections for the Higgs sector \cite{Heinemeyer:1998np,Degrassi:2001yf,Brignole:2001jy,Brignole:2002bz,Dedes:2003km}.  $\delta^{\rm SUSY} a_\mu$ was computed at full one-loop level adding the logarithmic piece of the quantum electro-dynamics two-loop calculation plus two-loop contributions from both stop-Higgs and chargino-stop/sbottom \cite{Heinemeyer:2004yq}. The effective two-loop effect due to a shift in the muon Yukawa coupling proportional to $\tan^2\beta$ has been added as well \cite{Marchetti:2008hw}.

Next we expound how a systematic analysis of this kind can be done with the help of Bayesian techniques. This will allow us to quantify the tension between $m_h$ and $a_\mu$ as a function of $m_h$.

\section{Quantifying the tension between $m_h$ and $a_\mu$}
\label{QuantifiyngTheTension}

Le us start by recalling some basic notions of Bayesian inference. We refer the reader to ~\cite{Trotta:2005ar,Trotta:2008qt} for further details. For a model defined by a set of parameters $\theta$, the posterior probability density function (pdf) of a point in parameter space, $\{\theta\}$, given a certain set of {\em data}, is denoted by $p(\theta|{\rm data})$ and it is obtained via Bayes theorem as
\bea
\label{Bayes}
p(\theta|{\rm data})\ =\  \frac{p({\rm data}|\theta)\ p(\theta)}{p({\rm data})}\ .
\eea
Here $p({\rm data}|\theta)$ is the likelihood function (when considered as a function of $\theta$ for the observed data)\footnote{Frequentist approaches, which are an alternative to the Bayesian framework, are based on the analysis of the likelihood function in the parameter space; see ref.~\cite{Buchmueller:2009fn} for a recent frequentist analysis of the MSSM.}. $p(\theta)$ is the prior, i.e. the probability density that we assign the points in the parameter space before seeing the data (in the context of Bayesian inference, the prior for a new cycle of observations can be taken to be the posterior from previous experiments). Finally, $p({\rm data})$ is a normalization factor, sometimes called the {\em evidence}. It is given by 
\bea
\label{BayesEvid}
p({\rm data})\ =\ \int d\theta  \  p({\rm data}|\theta)\ p(\theta)\ ,
\eea
i.e. the evidence is the average of the likelihood under the prior, and thus it gives the global probability of measuring the data in the model. 

When two different models (or hypotheses) are used to fit the data, the ratio of their evidences gives the relative probability of the two models in the light of the data (assuming equal prior probability for both). For an application to model selection in the context of the CMSSM, see \cite{Feroz:2008wr}.

In order to quantify the tension between $m_h$ and $a_\mu$, following Ref.~\cite{Feroz:2009dv} we separate the complete set of data in two subsets:
\bea
\label{DDD}
\{{\rm data}\}  = \{\mathscr{D}, D \}.
\eea
Here $\mathscr{D}$ represents the subset of observations, whose compatibity with the rest of the observations, ${D}$, (which are assumed to be correct) we want to test. In our case, $\mathscr{D}$ is the experimental value of $a_\mu$, whereas $D$ is given by all the standard electroweak observables, B- and D-physics observables, limits on supersymmetric masses, etc (for the complete list of experimental  data used in this paper, with references, see Table 2 of \cite{Cabrera:2009dm}). $D$ includes also the value of $m_h$ that we are probing, and thus provisionaly assumed to be the actual one. Hence, we will not consider any experimental error in the value of $m_h$, just the uncertainty associated to the theoretical calculation (estimated as $\pm 2$ GeV). Now we construct the quantity $p(\mathscr{D}| D)$, i.e. the probability of measuring a certain value for $\mathscr{D}$, given the known values of the remaining observables, $D$,
\bea
\label{pDD}
p(\mathscr{D}| D)\ =\ \frac{p(\mathscr{D}, D)}{p(D)}.
\eea
Here $p(\mathscr{D}, D)=p({\rm data})$ is the joint evidence, as given by Eq.~(\ref{BayesEvid}), i.e., the global probability of measuring both sets of data at the same time, and $p(D)$ is its equivalent but just for the $D$ subset. The latter is a normalization factor which will soon cancel out.

Now, the consistency of $\mathscr{D}^{\rm obs}$ (the measured muon anomaly) with the rest of data, $D$, in the context of the model (CMSSM), can be tested by comparing $p(\mathscr{D}^{\rm obs}|D)$ with the value obtained using different values of $\mathscr{D}$, in particular the one that maximizes such probability, say  $\mathscr{D}^{\rm max}$ (assuming the same reported error at the new central value). This
gives a measure of the likelihood of the actual data, $\mathscr{D}^{\rm obs}$, under the assumption that the model is correct:
\begin{equation}
\label{eq:Ltest}
\frac{p(\mathscr{D}^{\rm obs}|D)}{p(\mathscr{D}^{\rm max}|D)}\ =\ 
\frac{p(\mathscr{D}^{\rm obs},D)}{p(\mathscr{D}^{\rm max},D)}\
\equiv\ \ \mathscr{L} (\mathscr{D}^{\rm obs}|D). 
\end{equation}
$\mathscr{L} (\mathscr{D}^{\rm obs}|D)$ is analogous to a likelihood ratio in data space, but integrated over all possible values of the parameters of the model. Therefore, it can be used as a test statistics for the likelihood of the data being tested, $\mathscr{D}^{\rm obs}$, in the context of the model used (the CMSSM). Such test was called $\mathscr{L}-$test in Ref.~\cite{Feroz:2009dv}. Note that, as mentioned above, the $p(D)$ factor cancels out in the expression of $\mathscr{L} (\mathscr{D}^{\rm obs}|D)$, which is simply given by the ratio of the joint evidences. 

In our case, the value of $\mathscr{D}^{\rm max}$ depends on the value of $m_h$ probed. For very large $m_h$, say $m_h\geq 135$ GeV, SUSY must decouple, so $\mathscr{D}^{\rm max}$ should approach the SM prediction. Hence, in this limit one expects $\mathscr{L} (\mathscr{D}^{\rm obs}|D)$ to show a $~3.2\ \sigma$ discrepancy; in other words, $-2 \ln \mathscr{L} (\mathscr{D}^{\rm obs}|D)\rightarrow 3.2^2$. However, the expression \eqref{eq:Ltest} allows us to evaluate this likelihood for any intermediate value of $m_h$, and so we can evaluate how quickly this limit is reached as a function of the assumed value for $m_h$.

For the numerical calculation we have used the MultiNest \cite{Feroz:2007kg,Feroz:2008xx,Trotta:2008bp} algorithm as implemented in the \texttt{SuperBayeS} code \cite{superbayes,deAustri:2006pe,Roszkowski:2007fd}. It is based on the
framework of Nested Sampling, recently invented by Skilling \cite{SkillingNS,Skilling:2006}. MultiNest has been developed in such a way as to be an extremely efficient sampler even for likelihood functions defined over a parameter space of large dimensionality with a very complex structure as it is the case of the CMSSM. The main purpose of the Multinest is the computation of the Bayesian evidence and its uncertainty but it produces posterior inferences as a by--product at no extra computational cost. 

Fig.~1 shows the value of $-2 \ln \mathscr{L}$ (the analogous of the usual $\chi^2$) for different values of the Higgs mass,  $m_h ({\rm GeV})=115, 120,125, 130, 135$, and for two different choices of initial priors for the CMSSM parameters, namely log prior (red line) and flat prior (blue line). The precise shape of the log and flat priors used here is the one derived in ref.\cite{Cabrera:2009dm}, to which the reader is referred, which take into account the likelihood associated to the electroweak breaking process. The horizontal error bars reflect the uncertainty in the theoretical computation of $m_h$ in the MSSM, while the vertical error bars come from sources of error in the computation of $\mathscr{L}$, mainly the numerical accuracy of the evidence returned by MultiNest. Lines of conventional confidence levels thresholds in terms of number of $\sigma$ are shown as well for comparison.

\begin{figure}[t]
\begin{center}
\label{Ltest}
\includegraphics[width=1.0\linewidth]{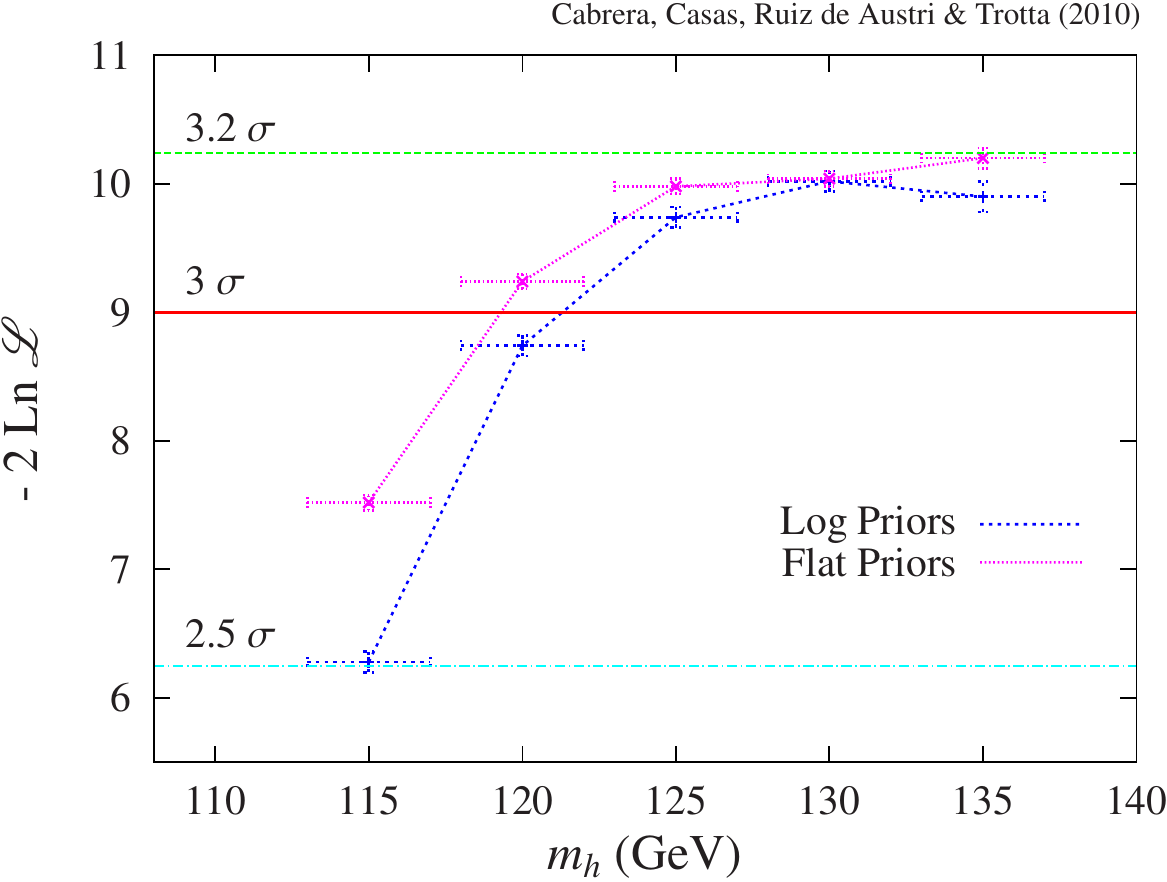}
\caption[text]{The $-2 \ln \mathscr{L}$ test statistics, as defined in Eq. (\ref{eq:Ltest}), as a function of the assumed value for $m_h$ in the CMSSM framework with logarithmic (blue, lower dotted line) and flat (violet, upper dotted line) priors. Horizontal lines denote thresholds of $2.5\ \sigma$, $3\ \sigma$ and $3.2\ \sigma$ discrepancy.} 
\end{center}
\end{figure}

From the figure we see that the likelihood of the experimental value of $a_\mu$ approaches asymptotically the expected 3.2~$\sigma$ discrepancy for large values of $m_h$, for both types of priors. As mentioned above, this is logical and it represents a nice cross-check of the reliability of the whole procedure. Besides, Fig.~1 tells us how fast this convergence is reached as $m_h$ increases. And, as a matter of fact, the convergence is very fast. At $m_h=125$ GeV the maximum level of discrepancy is already achieved, indicating that SUSY has decoupled, and thus the prediction for $a_\mu$ coincides with the SM one. If we require less than 3~$\sigma$ discrepancy, we need $m_h \lsim 120$ GeV. This is a prediction of the CMSSM provided we accept the calculation of $a_\mu$ based on $e^+e^-$ data. For a larger Higgs mass we should give up either the CMSSM model or the computation of $a_\mu$ based on $e^+e^-$; or accept living with such inconsistency. These are the main conclusions of this paper. They stem directly from Fig.~1. Let us also note that, even assuming a Higgs mass as low as it can be, the minimum level of discrepancy is about 2.5~$\sigma$. However, most of this tension with $a_\mu$ comes from $b\rightarrow s,\gamma$ data \cite{Feroz:2009dv}, rather from the value of the Higgs mass. This can be checked by repeating the analysis excluding all the experimental information (except $M_Z$ and the assumed Higgs mass). The resulting plot is similar to that of Fig.~1, except the $m_h=115$ GeV point, which shows a $\sim 1.5\ \sigma$ discrepancy.

It is an interesting exercise to compute how our conclusions would change if $a_\mu$ became more precisely measured in the future (keeping the same central value). If one continued to assume the theoretical evaluation of
$a_\mu$ based on $e^+e^-$ data, the signal for new physics would obviously become stronger. In this case, the tension between a large Higgs mass and the experimental $a_\mu$ would get more unbearable. We have done this excercise, by changing (artificially) the experimental uncertainty of $a_\mu^{\rm exp}$, so that the discrepancy with the SM result be $5\ \sigma$, something that could happen in the next years. Now, in the context of the CMSSM, the value of $-2 \ln \mathscr{L} (\mathscr{D}^{\rm obs}|D)$ must approach asymptotically such $5\ \sigma$ discrepancy, and this is indeed what we observe, as shown in Fig.~2. In this hypothetical situation, a Higgs mass above 120 Gev would imply a discrepancy larger than $4\ \sigma$ with the muon anomaly in the context of the CMSSM. Actually, the present lower bound, $m_h\geq 114.4$ GeV, would already be inconsistent with the muon anomaly at the $3\ \sigma$ level.

\begin{figure}[t]
\begin{center}
\label{Ltest5s}
\includegraphics[width=1.0\linewidth]{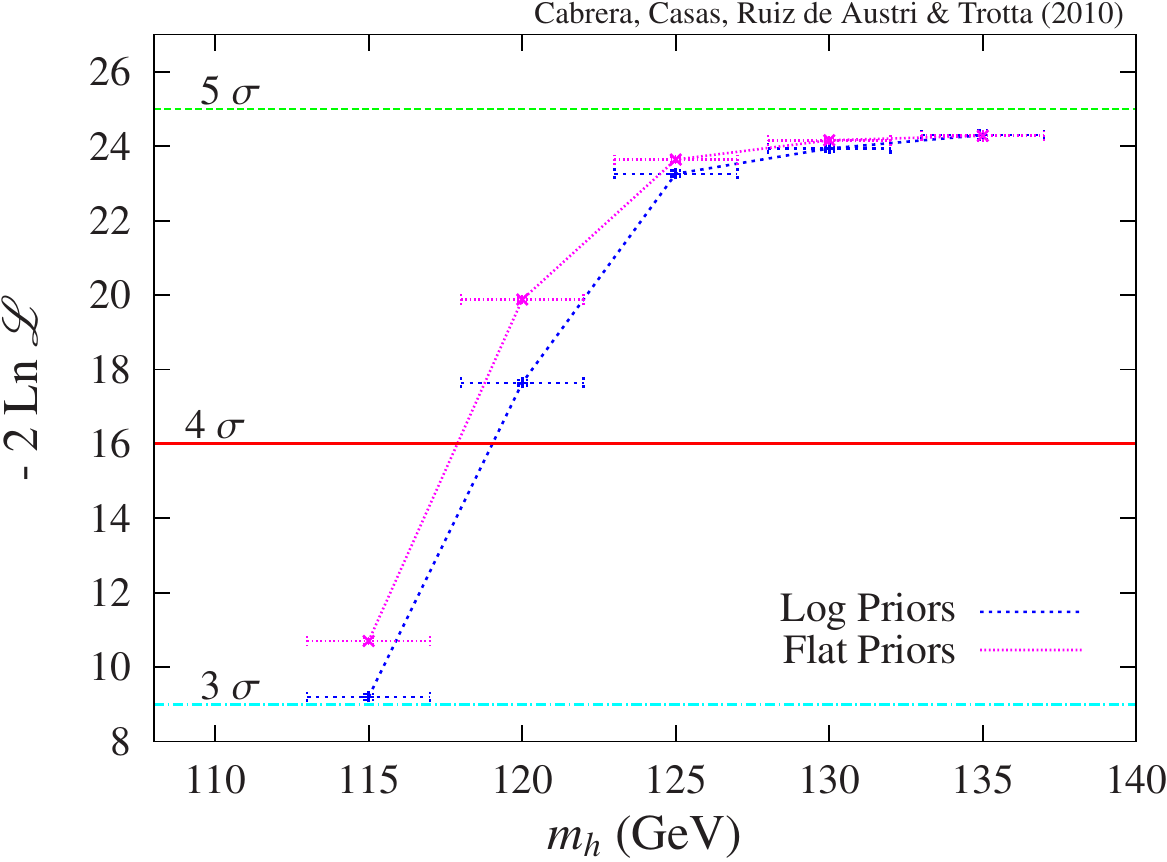}
\caption[text]{As in Fig.~\ref{Ltest}, but (artificially) assuming an improved experimental determination of $a_\mu$, so that the SM discrepancy becomes $5\ \sigma$. Now the horizontal lines denote thresholds of $3\ \sigma$, $4\ \sigma$ and $5\ \sigma$ discrepancy.} 
\end{center}
\end{figure}

This gives a fair idea of the tensions within the CMSSM to accommodate
a value of $a_\mu$ as the measured one (basing the theoretical
calculation on present $e^+e^-$ data).

\section{Probability distributions for supersymmetric parameters}
\label{pdfs}

It is also interesting to investigate the probability distributions of the CMSSM parameters for various assumed values of the Higgs mass. Figure 3 (upper panels) shows the marginalized probability distribution functions (pdfs) of $m$, $M$
assuming a value of $m_h=115, 120, 125$ (GeV), as well as adding in all present-day constraints mentioned above. The location of the peak in the posterior pdf increases with the assumed Higss mass since, as mentioned in section \ref{higgs:gm2}, in the MSSM a large $m_h$ requires large radiative contributions, which grow logarithmically with the stops masses. This happens
even though large values of $m$ and $M$ are penalized both for a natural electroweak breaking (see ref. \cite{Cabrera:2008tj,Cabrera:2009dm}) and by the need of a sizeable  $\delta^{\rm SUSY}a_\mu$. The model ``prefers" to reproduce $m_h$ at the cost of not reproducing $a_\mu$ rather than viceversa. Note here that for increasing soft masses the discrepancy of $a_\mu$ with the experimental value approaches $3.2\ \sigma$, but if the soft masses are not large enough, the discrepancy associated to $m_h$ would be much more severe.

\begin{figure*}[t]
\begin{center}
\label{m0:m12:tanb}
\includegraphics[angle=0,width=0.35\linewidth]{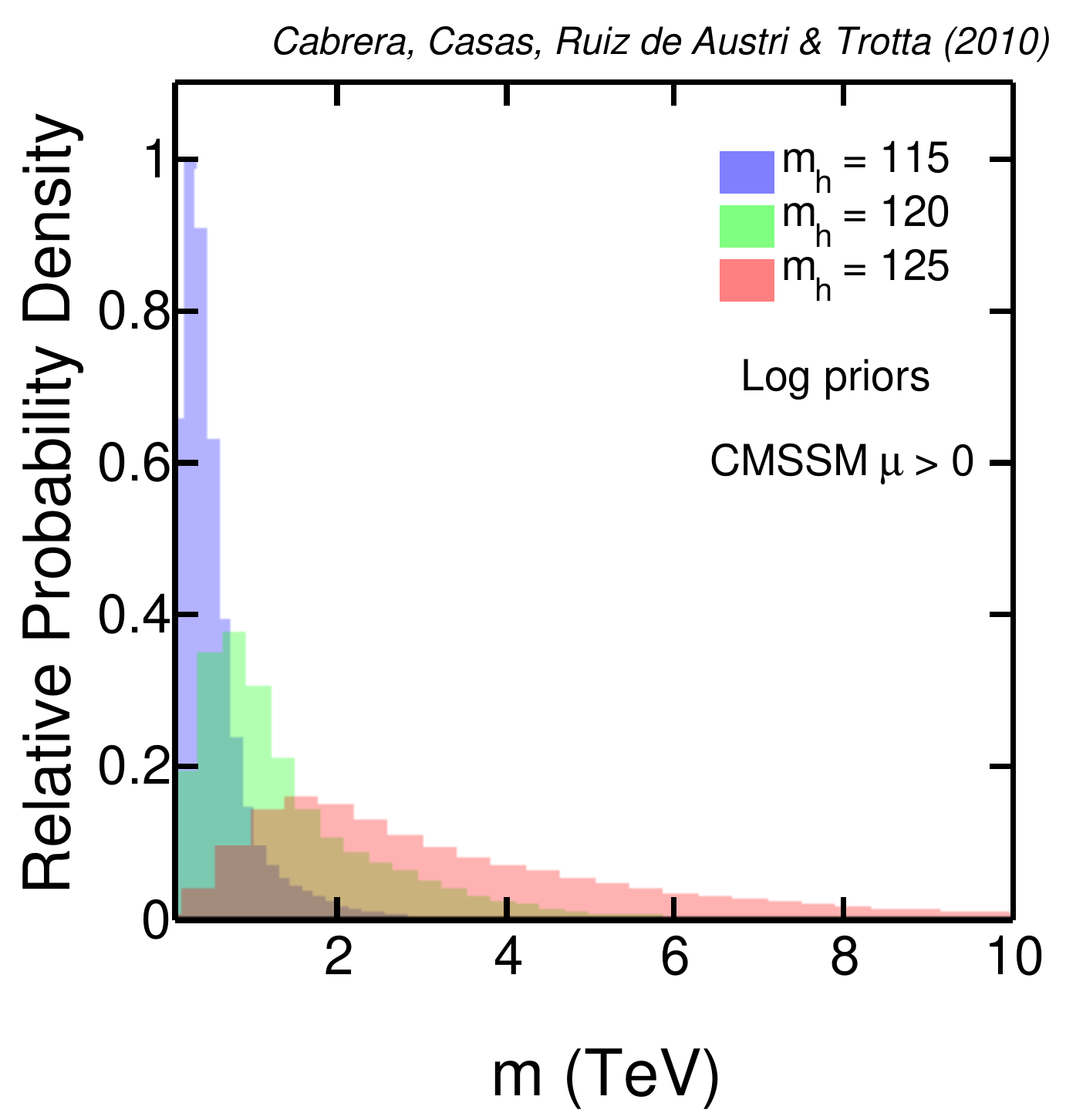} \hspace{0.7cm}
\includegraphics[angle=0,width=0.35\linewidth]{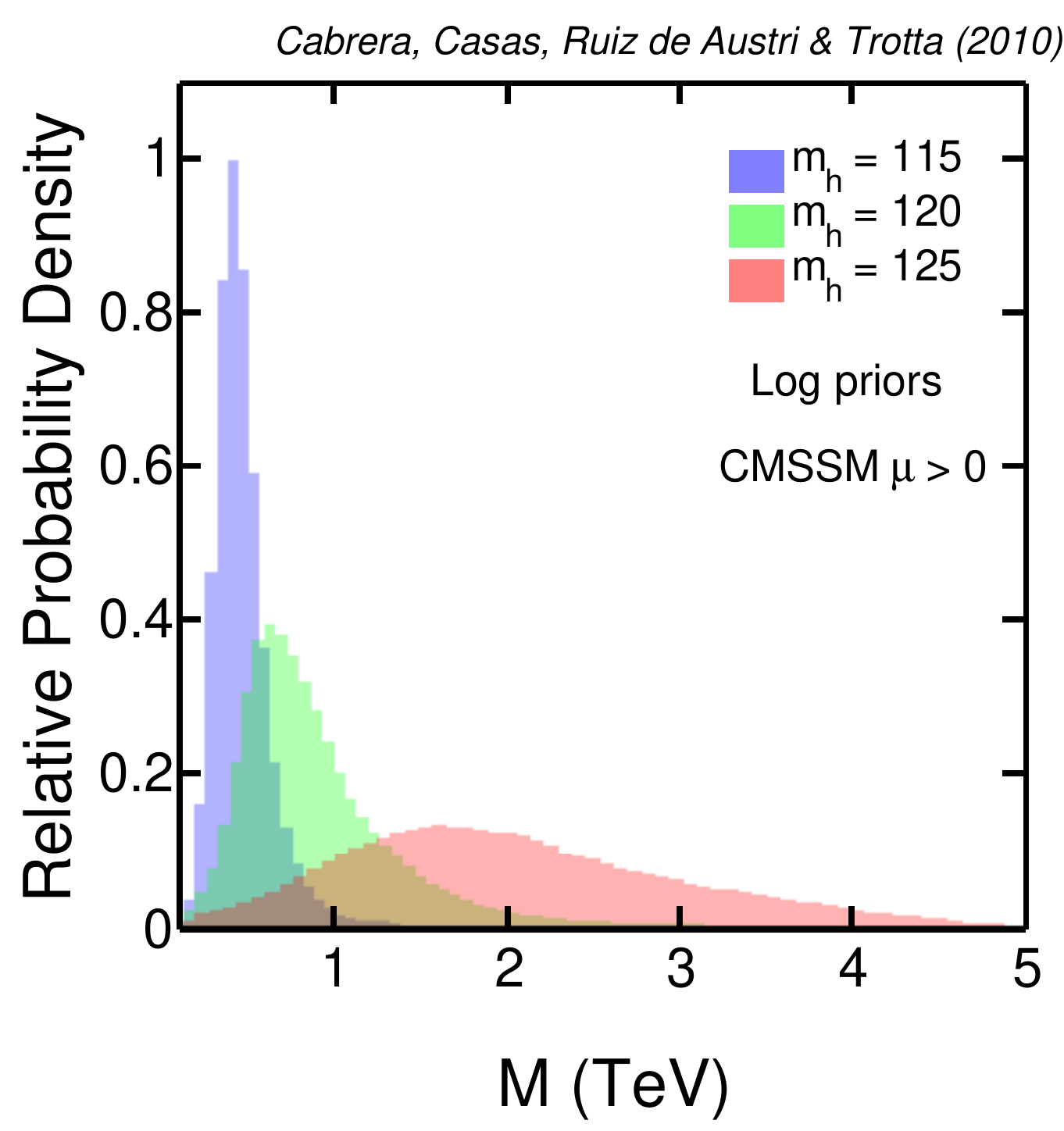} \\ \vspace{0.5cm}
\includegraphics[angle=0,width=0.35\linewidth]{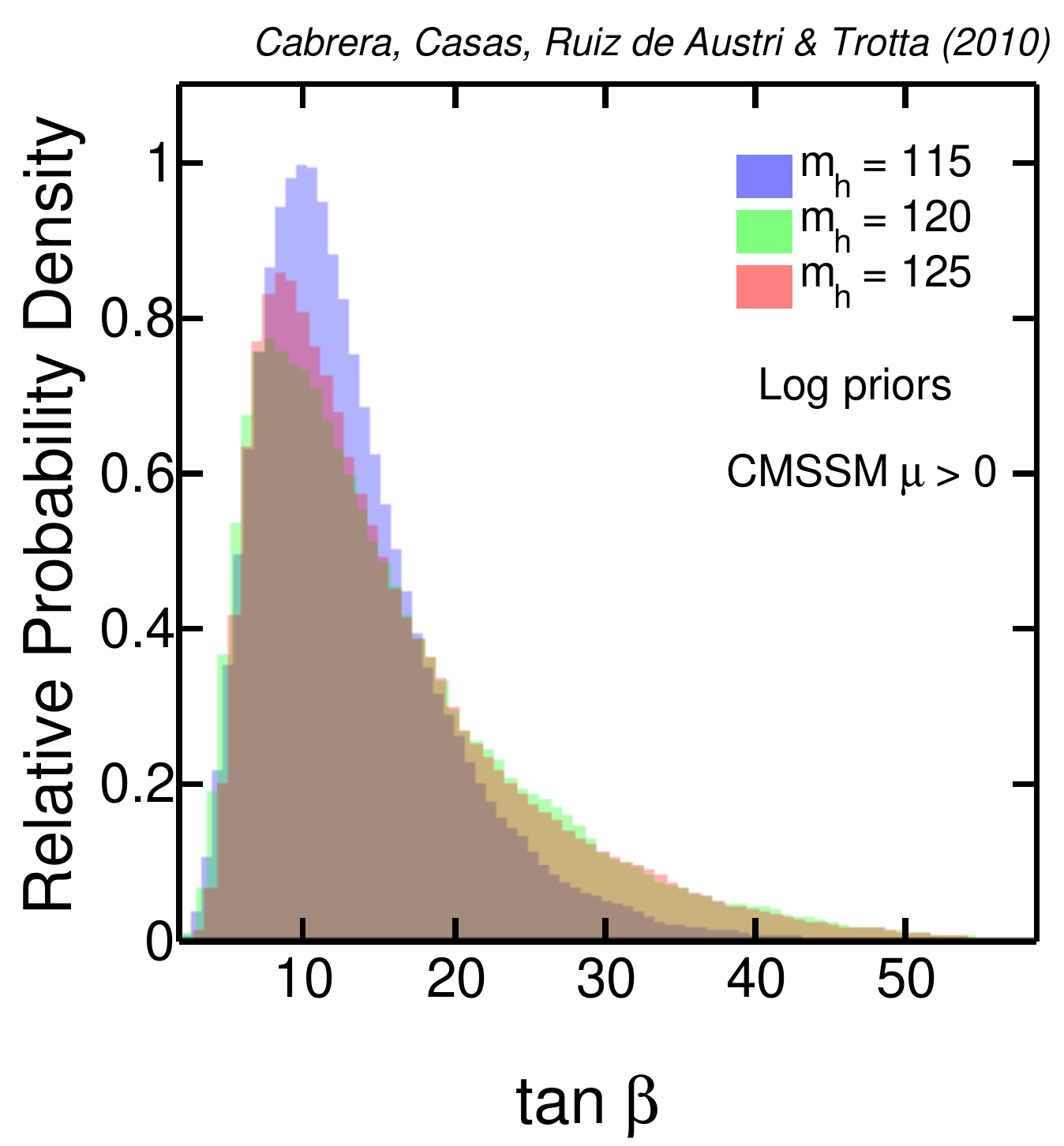}
\caption{Probability distribution functions of $m$, $M$ (upper
  panels) and $\tan\beta$ (lower panel) for $m_h=$ 115 GeV (blue), 120
  GeV (green) and 125 GeV (red).}
\end{center}
\end{figure*}

Fig.~3 (lower panel) shows the pdf of $\tan\beta$ for $m_h=115, 120, 125$ (GeV). Its shape is the result two competing effects. On the one hand, large values of $\tan\beta$ are severely penalized for the electroweak breaking \cite{Cabrera:2008tj,Cabrera:2009dm}. On the other hand, the need of a sizeable $\delta^{\rm SUSY}a_\mu$ favours large $\tan\beta$ (see the approximate expression (\ref{g-2})). Fig.~3 shows the balance between these two effects. [The Higgs mass increases also with $\tan\beta$, but the effect is only important for small values of $\tan\beta$, see eq.(\ref{mhMSSM})]. Now, since for larger $m_h$ the soft masses are larger, with the side-effect of suppressing $\delta^{\rm SUSY}a_\mu$, one might expect that the preferred value of $\tan\beta$ increases with $m_h$, to compensate this in eq.~(\ref{g-2}). However, this effect is not very important, as it is apparent in Fig.~3.
To understand this, let us approximate (for the sake of the argument) $M_{susy}\sim m_{\tilde{e}_{L,R}}$ in eq.~(\ref{g-2}) and use \cite{Martin:1997ns}, \cite{Casas:2003jx}
\bea
\label{sparticles}
m_{\tilde{e}_L}^2 &\simeq& m^2 + 0.54 M^2,\\ \nonumber
m_{\tilde{e}_R}^2 &\simeq& m^2 + 0.15 M^2,\\ \nonumber
m_{\tilde{t}}^2 &\simeq& 3.36M^2 + 0.49m^2 - 0.05A^2 - 0.19AM + m_t^2 .
\eea
Since $m_h$ increases (logarithmically) with $m_{\tilde{t}}^2$, while $\delta^{\rm SUSY}a_\mu$ is suppressed by $m_{\tilde{e}_{L,R}}^2$, it might seem that the most efficient way to reproduce both is to increase $M$ rather than $m$ (note the different dependences on $M$ in eqs.~(\ref{sparticles})). The problem is that the fine-tuning grows very fast with $M$; in other words, the number of points in the parameter space with correct EW breaking decresases very quickly. In consequence this possibility is statistically penalized.
On the contrary, for small $M$ and large $m$, if $\tan\beta>8$, there is a focus-point region, with small fine-tuning. This region is statistically favoured, though this is counteracted by the penalization arising from the suppression in $\delta^{\rm SUSY}a_\mu$. This cannot be compensated by larger values of $\tan\beta$, since in this regime very big values of $\tan\beta$ (as would be needed for such compensation) start to be forbidden as we increase $m$. In consequence, a very large $\tan\beta$ is hardly favoured by an increasing $m_h$.

Finally, let us mention that a lot of effort has been done in the literature to determine the most probable region of the parameter space of the CMSSM  \cite{Buchmueller:2009fn,AbdusSalam:2009qd,Ellis:2008di,Trotta:2008bp,Buchmueller:2008qe,Allanach:2007qk,deAustri:2006pe,Allanach:2006jc,Allanach:2005kz,Beskidt:2010va,Buchmueller:2010ai,Cabrera:2009dm,Cabrera:2008tj}. This includes both Bayesian approaches (as the one followed here) and frequentist ones. The latter (which can be considered as complementary to the Bayesian ones) are based on the analysis of the likelihood function in the parameter space. Thus they do not  penalize regions from fine-tuning arguments (something automatic in Bayesian analyses \cite{Cabrera:2008tj,Cabrera:2009dm} ). In consequence, following a frequentist approach it would be much more hard to show up the tension between $m_h$ and $g-2$. On the other hand, the present analysis differs from the previous ones in the fact that several hypothetic future scenarios, depending on the value of the Higgs mass, are considered and compared.

\section{Conclusions}

As it is well known, the SM prediction for the magnetic anomaly of the muon, $a_\mu$ (basing the computation of the hadronic contribution on $e^+e^-$ data) shows a $>3\ \sigma$ discrepancy with the experimental result. It is common to consider 
this discrepancy as a signal of new physics (though, admittedly, the theoretical computation is controversial). In that case, SUSY is a most natural option for such new physics. 

However, as we have discussed in this paper, in the supersymmetric context there is a potential tension between the requirement of SUSY contributions to the muon anomaly, $\delta^{\rm SUSY}a_\mu $, sufficient to reconcile theory and experiment, and a possibly large Higgs mass. In the CMSSM framework a large Higgs mass means ${\cal O}(10)$ GeV above the present experimental bound, $m_h\geq 114.4$ GeV (for an SM-like Higgs). The tension arises because the main contributions to $\delta^{\rm MSSM}a_\mu $ come from 1-loop diagrams with chargino-sneutrino and neutralino-smuon exchange, which grow with decreasing supersymmetric masses and increasing $\tan\beta$. But, on the other hand, in the MSSM the tree-level Higgs mass is bounded from above by $M_Z$, so radiative corrections (which grow logarithmically with the stop masses) are needed to reconcile the theoretical predictions with the present experimental lower bound. Thus, a large Higgs mass requires large supersymmetric masses, making impossible the task of reproducing  the experimental value of $a_\mu$.

Although it is clear that in the limit of very large $m_h$ (say above 135 GeV) the CMSSM must present the same discrepancy as the SM regarding the prediction for $a_\mu$, it is much less clear for which size of $m_h$ does the tension start to be unbearable, and would therefore put the model under pressure. Note that a particular value of the Higgs mass, say $m_h = 120$ GeV, can be achieved through very different combinations of the supersymmetric parameters, producing different values of $\delta^{\rm MSSM}a_\mu $. On the other hand, it may happen that, for a given value of $m_h$, the region of the parameter space compatible with $a_\mu^{\rm exp}$ is extremely tiny, implying a fine-tuning and thus a tension between the two observables.

Our goal has been to quantify such tension, as a function of $m_h$, with the help of Bayesian techniques. As discussed at the end of sec. IV, this is the natural approach if we want to incorporate the statistical penalization of fine-tuned regions of parameter-space. Certainly, if one just assumed a particular supersymmetric model (i.e. a point in the parameter-space, no matter how fine-tuned it were) then the statistical arguments used in this paper would not be appropriate. We have shown that for $m_h\geq 125$ GeV the maximum level of discrepancy ($\sim 3.2\ \sigma$) is already achieved, indicating that SUSY has decoupled, and thus the prediction for $a_\mu$ coincides with the SM one. Given present-day data, requiring less than a 3~$\sigma$ discrepancy, implies $m_h \lsim 120$ GeV. This is a prediction of the CMSSM provided we accept the calculation of $a_\mu$ based on $e^+e^-$ data. For a larger Higgs mass we should give up either the CMSSM model (a the 3~$\sigma$ level or above) or the computation of $a_\mu$ based on $e^+e^-$; or else accept living with such inconsistency. These are the main conclusions of this paper, and can be inferred directly from Fig.~1. It is also important to note that, as discussed in section \ref{QuantifiyngTheTension}, the CMSSM cannot remove the full $3.2 \sigma$ discrepancy in $a_{\mu}$.

We have also examined the possibility that the experimental uncertainty of $a_\mu^{\rm exp}$ will decrease in the future, so that the discrepancy with the SM result be $5\ \sigma$, something that could happen in the next years. Then, in the context of the CMSSM, a Higgs mass above 120 GeV would imply a discrepancy larger than $4\ \sigma$ with the muon anomaly. Actually, the present lower bound, $m_h\geq 114.4$ GeV, would already be inconsistent with the muon anomaly at the $3\ \sigma$ level. This illustrates the tensions within the CMSSM to accommodate a value of $a_\mu$ as the measured one (basing the theoretical calculation on present $e^+e^-$ data).

Finally, we have shown how the probability distributions of the most relevant parameters (universal scalar and gaugino masses, and $\tan\beta$) change with increasing $m_h$, which has obvious implications for the detection (or non-detection) of SUSY in the LHC.

\section*{Acknowledgements}

This work has been
partially supported by the MICINN, Spain, under contracts
FPA-2007--60252 and FPA-2007-60323; Consolider-Ingenio PAU
CSD2007-00060 and MULTIDARK CSD2009-00064. We thank as well the
Generalitat Valenciana grant PROMETEO/2008/069;
the Comunidad de Madrid through Proyecto HEPHACOS ESP-1473 and the
European Commission under contract PITN-GA-2009-237920.%
R.T. would like to thank the EU FP6 Marie Curie Research and Training
Network ``UniverseNet'' (MRTN-CT-2006-035863) for partial support and
the U.A.M. for hospitality. %
M.~E. Cabrera acknowledges the financial support of the CSIC through a
predoctoral research grant (JAEPre 07 00020).%
The use of the ciclope and hydra cluster of the IFT-UAM/CSIC is also
acknowledged.

\bibliographystyle{JHEP-2}
\bibliography{references}

\end{document}